\documentclass{article}
\usepackage[T1]{fontenc} 
\usepackage[utf8]{inputenc} 
\usepackage{ismir,amsmath,cite,url}
\usepackage{graphicx}
\usepackage{color}
\usepackage{tabularx} 

\usepackage{lineno}
\usepackage{booktabs}
\usepackage{arydshln}
\usepackage[normalem]{ulem}
\usepackage{todonotes}

\title{Robust lossy audio compression identification}

\multauthor
{Hendrik Vincent Koops \hspace{1cm} Gianluca Micchi \hspace{1cm} Elio Quinton} { 
Music and Audio Machine Learning Lab, \\
Universal Music Group, London, UK \\
{\tt\small \{vincent.koops, gianluca.micchi, elio.quinton\}@umusic.com }
}

\def\authorname{F. Author, S. Author, and T. Author}

\begin{document}

\newcommand{\mpthree}{\textsc{mp\oldstylenums{3}}}

\maketitle
\begin{abstract}
Previous research contributions on blind lossy compression identification report
near perfect performance metrics on their test set, across a variety of codecs
and bit rates. However, we show that such results can be deceptive and may not
accurately represent true ability of the system to tackle the task at hand. In
this article, we present an investigation into the robustness and generalisation
capability of a lossy audio identification model. Our contributions are as
follows. (1) We show the lack of robustness to codec parameter variations of a
model equivalent to prior art. In particular, when naively training a lossy
compression detection model on a dataset of music recordings processed with a
range of codecs and their lossless counterparts, we obtain near perfect
performance metrics on the held-out test set, but severely degraded performance
on lossy tracks produced with codec parameters not seen in training. (2) We
propose and show the effectiveness of an improved training strategy to
significantly increase the robustness and generalisation capability of the model
beyond codec configurations seen during training. Namely we apply a random mask
to the input spectrogram to encourage the model not to rely solely on the
training set's codec cutoff frequency. 


\end{abstract}
\section{Introduction}\label{sec:introduction}

Audio codecs can be roughly categorized into two categories: \textit{lossless}
and \textit{lossy}. \textit{Lossless} means that an exact preservation of the
signal is guaranteed by the codec. In other words, the signal resulting from
encoding and decoding is exactly identical to the original.  
In contrast, \textit{lossy} encoding means that some of the signal is lost in
the encoding and decoding process. In other words, the signal resulting from
encoding and decoding is not exactly identical to the original signal.

Popular lossy audio codecs like \mpthree{} \cite{musmann2006genesis},  Ogg
Vorbis \cite{vorbis} or \textsc{aac} \cite{bosi1997iso} are known as
"perceptual" codecs because they rely on models of human auditory cognition to
prioritise the deletion of parts of the audio signal that have the least
perceptual impact on human listeners. Despite the signal degradation that they
result in, perceptual lossy codecs can achieve much greater compression ratios
than lossless codecs, and are therefore well suited for applications where data
bandwidth is limited. For example, they have been instrumental in enabling music
streaming over networks with limited bandwidth. 
 
Digital audio codecs are readily available and are integrated into many
widespread professional and consumer tools such as Digital Audio Workstations,
software libraries, digital music players etc., which make converting an audio
file from one format to another nowadays extremely easy and accessible to
anyone. As a result it is easy to mistakenly encode a source audio signal with a
lossy codec, which degrades the signal, and then decode it back into a lossless
file container. This process may create the illusion that a lossless file
container (e.g.\ \textsc{wav}) contains unimpaired audio when it does in fact
contain lossy-compressed audio. 

Guaranteeing audio integrity is essential in many applied scenarios such as
large scale music distribution or archiving. Because the aforementioned case of
lossy audio disguised as a lossless file would violate this guarantee, there is
a need to automatically detect such occurrences. Identification of audio that
has been compressed with a lossy codec is a valuable component of quality
assurance processes, which form an important part of many modern musical audio
content pipelines. 

\textbf{Contributions.} In this paper, we present an investigation into the
robustness and generalisation capability of a lossy audio identification model.
We show that when we naively train a lossy compression detection model on a
dataset of music recordings processed with a range of codecs and their lossless
counterparts, we obtain near perfect performance metrics on the held-out test
set. However, we obtain severely degraded performance on lossy tracks produced
with codec parameters not seen in training. We also propose a new training
schema in which we randomly mask the input spectrogram to improve the model's
robustness. We show that our approach significantly increases the robustness and
generalisation capability of the model beyond codec configurations seen during
training.


\section{Background}

In the following sections, we will first provide a high level overview of lossy
audio codecs (Section \ref{sec:codecs}). Next, in Section \ref{sec:lossy_refs},
related work on lossy audio identification is discussed. Finally, in Section
\ref{sec:robustness}, we briefly present related work in MIR on robustness
evaluation.

\subsection{Lossy Codecs}
\label{sec:codecs}
\begin{figure}
    \centering
    \includegraphics[trim={0cm 10cm 16cm 1cm}, width=\columnwidth]{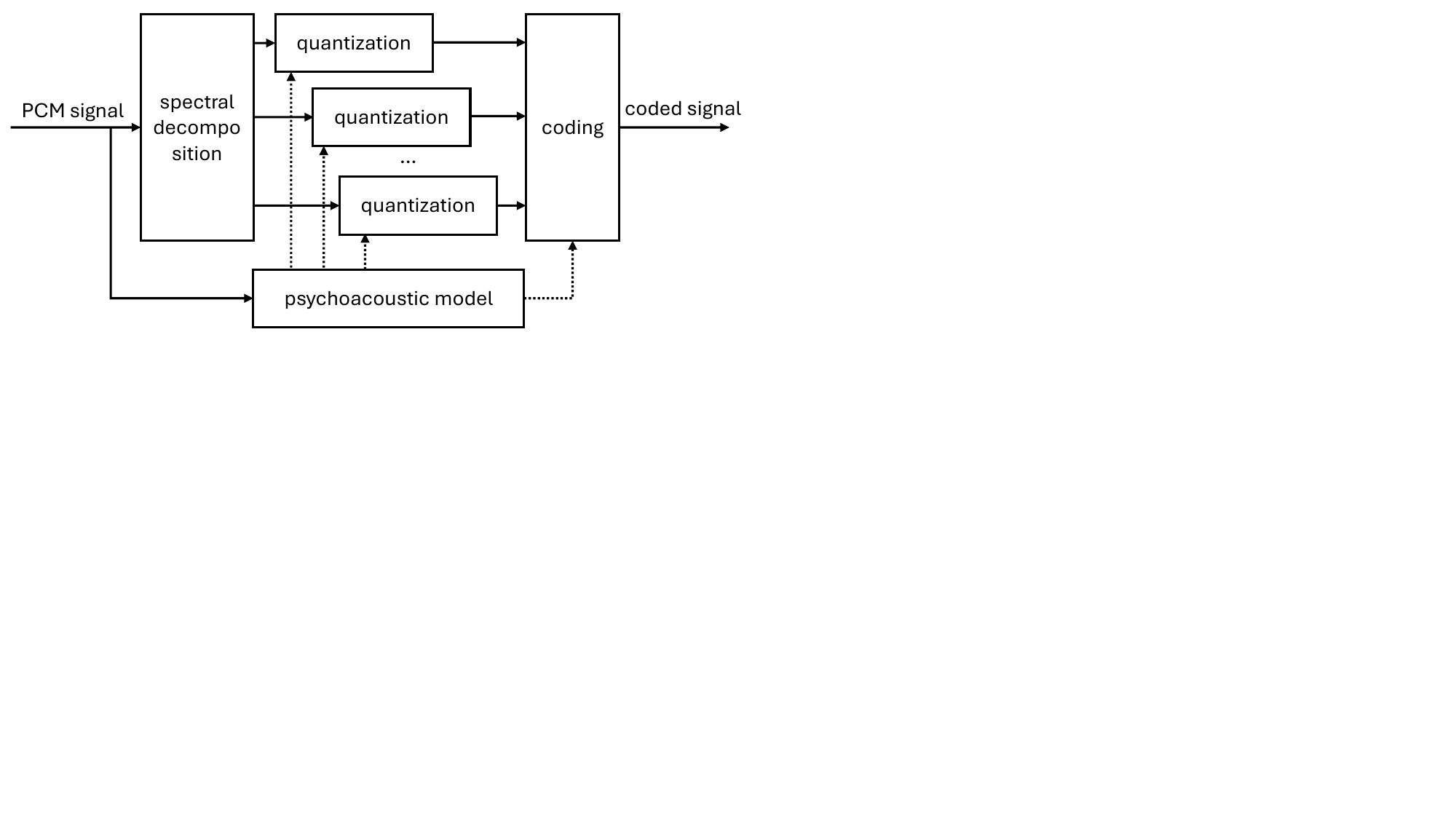}
    \caption{Basic block diagram of a perceptual audio coder. After spectral
    decomposition, a psychoacoustic model informs the quantization of individual
    spectral components.}
    \label{fig:coder}
\end{figure}

Figure \ref{fig:coder} shows a basic block diagram with the common modules of a
perceptual audio coder. The process of encoding an audio signal with a lossy
codec is commonly as follows. First, the original uncompressed (often pulse code
modulated - \textsc{pcm}) signal is transformed into a time-frequency
representation. This is typically done using a modified discrete cosine
transform (\textsc{mdct}), but many other transforms have been proposed
\cite{bosi2002introduction}. Commonly used signal block for the spectral
decomposition are between 2ms and 50ms. The components of the spectral
decomposition are then individually quantized. The quantization of the spectral
components is controlled by a psychoacoustic model that describes the time and
frequency masking properties of the human auditory system. Auditory masking is a
process where one sound (maskee) becomes inaudible in the presence of another
sound (mask) \cite{gelfand2017hearing}.

\begin{figure}
    \centering
    \includegraphics[width=1\columnwidth, trim={0cm 13cm 21cm 0cm}]{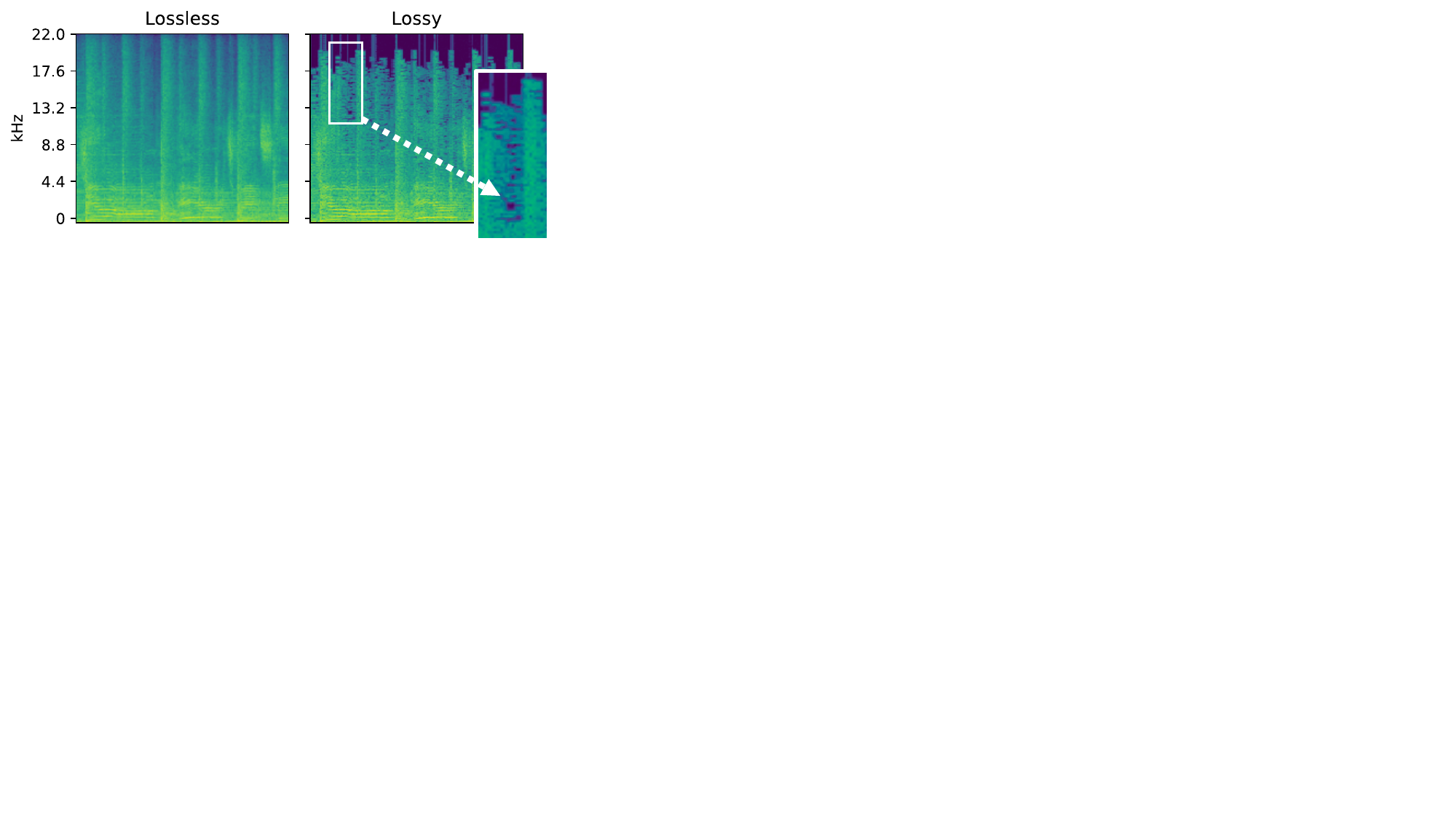}
    \caption{Spectrograms of examples of a lossless (left) and lossy version of
    the same audio excerpt (right). The latter is compressed with the
    \textsc{libfdk\_aac} codec at 128 kbps bit rate. The version on the right
    shows the hallmarks of lossy compression:
    removal of \textsc{fft} coefficients, holes in the spectrum, and general
    loss of higher frequency content. }
    \label{fig:lossyvslossless}
\end{figure}

Auditory masking can occur in the time domain (temporal masking) or in the
frequency domain (frequency masking). The quantization controlled by the
psychoacoustic model effectively controls which spectral coefficients will be
removed, resulting in spectral band rupture and holes in spectrograms, as
observed in Figure \ref{fig:lossyvslossless}. After quantization, Huffman coding
(or some other form of entropy coding) is applied to remove or reduce the
redundancy in the signal \cite{huffman1952method}. The bit rate of a codec
effectively controls both the size and the perceptual quality of the audio. A
low bit rate (like 128 kbps) will produce a small storage footprint, but
generally worse perceptual quality compared to a higher bit rate (like 320
kbps). For more detail on audio codecs and standards, we refer to
\cite{bosi2002introduction}.

\subsection{Lossy Compression Identification}
\label{sec:lossy_refs}

In previous research, multiple blind lossy compression identification models
have been proposed. These can broadly be categorized into two approaches. One
approach is to estimate codec parameters from the audio signal, to determine
factors such as the decoder framing grid, filter bank parameters and/or
quantization information. This type of approach has been successfully applied for
individual codecs like AAC \cite{herre2000analysis}, \mpthree{}
\cite{biessmann2013estimating, moehrs2002analysing, hiccsonmez2011audio}.
Although this type of approach can be very effective, it is computationally very
expensive, especially when multiple codecs are considered.

The second method utilizes audio quality measures to determine whether the audio
is lossy. One effect of lossy audio compression is the introduction of ``holes''
in the spectrogram, especially right after louder transients. This is the result
of the fact that spectral coefficients can be removed when they are perceptually
masked by other coefficients. Therefore, most approaches present some form of
``hole-detection``, such as estimating the number of inactive spectral
coefficients (e.g. \cite{moehrs2002analysing, yang2009defeating}) or computing
spectral fluctuations \cite{yang2008detecting, kim2018lossy,
derrien2019detection, d2009mp3}.

In \cite{hennequin2017codec}, Hennequin et al. presented a method for detecting
lossy compression based on a convolutional neural network \textsc{cnn} 
applied to audio spectrograms. 
Similarly, Seichter et al. in \cite{seichter2016aac} also proposed
a \textsc{cnn} approach for \textsc{aac} encoding detection and bit rate
estimation. All research contributions on lossy compression identification
almost uniformly report near-perfect performance metrics on their test set,
across a variety of codecs and bit rates. 

However, most codecs can be configured with parameters other than the bitrate
too, such as a cutoff frequency that controls the amount of higher frequencies
that will be preserved. \textsc{aac} for example has a default cutoff frequency
of around 17kHz \cite{aac_codec} for constant bit rates of 96 kbps per channel
and above, which means that the bandwidth of the encoder is set to 0 - 17kHz.
None of the previous research explores what happens when this parameter is
changed. 

In this paper we show that a model naively trained on default parameters may not
efficiently learn to discriminate lossy audio encoded with different parameters
and we analyse what happens when varying the cutoff frequency as an example.
Therefore, the good results previously reported must be taken with a pinch of
salt.

\subsection{Robustness Evaluation in Music Information Retrieval}
\label{sec:robustness}

Several studies in music information retrieval have shown that models can
seemingly achieve very high evaluation performance, while further research
reveals that what those models have learned is some confound with the ground
truth dataset \cite{kereliuk2015deep}. For example, in a research into the
robustness of genre classification models, Sturm showed that although these
systems might have high mean classification accuracies, they don't actually
reflect the underlying properties of the genre \cite{sturm2012two}. Furthermore,
it is shown that by filtering the audio signal in a minimal way, the models
produce radically different genre predictions. For a larger overview of music
adversaries in music information retrieval research, we refer to
\cite{kereliuk2015deep}. Bob Sturm in \cite{6847693} introduced the term
``horse''\footnote{A nod to the Clever Hans horse, see
\url{https://en.wikipedia.org/wiki/Clever_Hans}} to refer to system appearing
capable of achieving high evaluation performance, but actually working by using
irrelevant characteristics (confounds), and therefore not actually addressing
the problem it appears to be solving.

\section{Method}

In the following sections, we will first describe our model setup (in Section
\ref{sec:arch}), then our dataset (in Section \ref{sec:datasets}) 
and finally our proposed evaluation methods (in Section \ref{sec:eval}).

\subsection{Network Architecture}
\label{sec:arch}

\begin{figure}
    \centering
    \includegraphics[trim={0cm 11cm 9cm 2.8cm},clip,width=2\columnwidth]{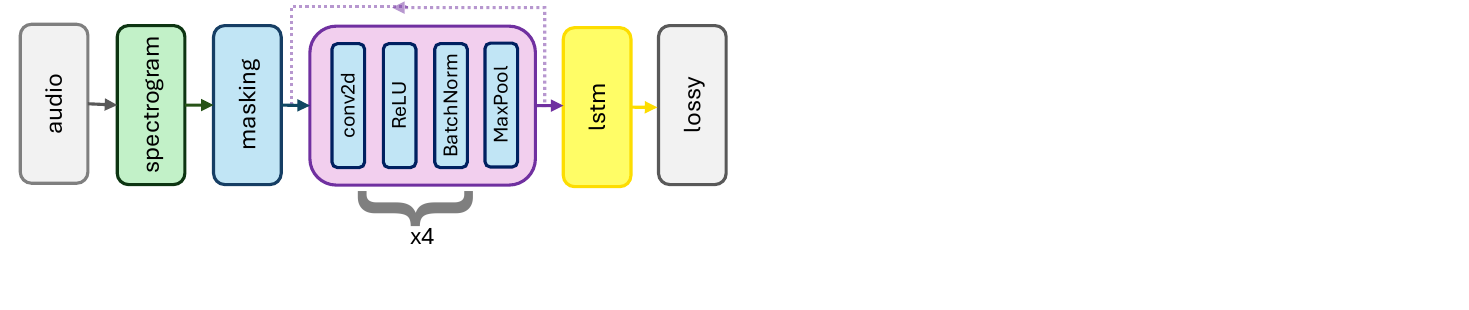}
    \caption{Proposed model for the detection of lossy audio Our model takes as
    input 2 seconds of audio, which is passed to a torchaudio spectrogram layer
    (in green). Depending on the experiment, the spectrogram is then passed to a
    masking layer (in blue), which  simulates low-pass filtering. The
    spectrogram is then passed to four convolutional modules (in pink). We use a
    bi-directional \textsc{lstm} (in yellow) for dimensionality reduction.
    We classify the audio into lossy or lossless in the final model head.
    }
    \label{fig:model}
\end{figure}

For the detection of lossy audio we propose a model (visualized in Figure
\ref{fig:model}) that can be divided into four parts: a spectrogram + random
mask module, 4 convolutional blocks, an \texttt{lstm} block and a classification
head made of a single dense layer. The architecture is partly inspired by prior
work by Hennequin et al. in \cite{hennequin2017codec} and Seichter et al. in
\cite{seichter2016aac}. In the following sections, we will describe each part in
detail. 

The model takes as input 2 seconds of raw monophonic audio signal sampled at
44.1 kHz, which is passed to a \emph{torchaudio} spectrogram layer that produces 
a magnitude spectrogram with 1024 \textsc{fft} coefficients \cite{hwang2023torchaudio}.

\textbf{Random mask.}
A random mask is optionally applied to the input spectrogram. This is achieved
by uniformly randomly sampling a cutoff frequency between 14 kHz and the Nyquist
frequency of the sample, and nulling all \textit{fft} coefficients above that
frequency by setting them to the minimum of the input spectrogram. A similar
approach called Specaugment was proposed by Park et al. in
\cite{park2019specaugment}.

In our first experiment (as described in Section \ref{sec:experiment1}) this
layer is not used, and the spectrogram is directly fed to the convolutional
blocks. However, in the second experiment (as described in Section
\ref{sec:experiment2}), we use this random mask layer with a different random
cutoff frequency for every training example.

\textbf{CNN.}
Each of the CNN blocks consist of four layers: a 2D convolutional layer with a
kernel size of (3,3), a ReLU layer, a batch normalization layer and a 2D
max-pooling layer. The max pooling size for each block is (2,2), with the
exception of the last block, which is (2,4).

\textbf{LSTM.}
We connect the \textsc{cnn} to a \emph{long short-term memory} (\textsc{lstm})
block for two reasons. Firstly, we want to exploit possible sequential
properties of the CNN output, and secondly, for dimensionality reduction for the
last (dense) part of the network. We use a bidirectional \textsc{lstm} with two
layers of size 128. 

\textbf{Classification head.}
Our model's lossy/lossless classification head is connected to the \textsc{lstm}
output with a dense layer of size 256 (2x 128 because our \textsc{lstm} is
bi-directional). The classification head has a softmax activation and 2 outputs
that model the probability of the example being lossless or lossy.

\textbf{Training.} 
We back-propagate our model on the binary cross entropy of the classification
head and the ground truth. For each audio track, we take a 2-second random crop
at training time. 

\subsection{Datasets}
\label{sec:datasets}

For our experiments, we sample 10k tracks of lossless 16 bit, 44.1kHz
\textsc{wav} files from a large private library of commercial music. From these
tracks, we create two datasets.

\subsubsection{\textsc{ds\oldstylenums{1}}.} For the first dataset we encode
each track with a codec randomly chosen among \textsc{libmp3lame} (\mpthree{}),
\textsc{libfdk\_aac} (\textsc{aac}) and \textsc{libvorbis} (\textsc{ogg}), with
bit rate also randomly chosen among 128, 256 and 320 kbps.
Each encoded file is then decoded back into a 16bit, 44.1 kHz \textsc{wav} file
that is used as input to the model. All the encoding/decoding is done using
\textit{ffmpeg} \cite{ffmpeg}. Between lossless and lossy tracks, the dataset
comprises of 20k tracks.

\begin{table*}[t]
    \centering
        \begin{tabular}{l|rrrrrrrrr|r|r}
        Codec & 
        \multicolumn{3}{c}{\textsc{libfdk\_aac}} &
        \multicolumn{3}{c}{\textsc{libvorbis}} &
        \multicolumn{3}{c}{\textsc{libmp3lame}} & 
        Lossless & Mean \\
        
        Bit rate & 
        \rotatebox[origin=l]{0}{128k} & 
        \rotatebox[origin=l]{0}{256k} &
        \rotatebox[origin=l]{0}{320k} &

        \rotatebox[origin=l]{0}{128k} & 
        \rotatebox[origin=l]{0}{256k} &
        \rotatebox[origin=l]{0}{320k} &

        \rotatebox[origin=l]{0}{128k} & 
        \rotatebox[origin=l]{0}{256k} &
        \rotatebox[origin=l]{0}{320k} &
        --- & --- \\
        
        \hline
        \textsc{ds\oldstylenums{1}} & 100.0 & 98.91 & 100.0 & 100.0 & 100.0 &
        100.0 & 100.0 & 100.0 & 98.37 & 99.88 & 99.79 \\
        \textsc{ds\oldstylenums{2}} & 31.38 & 28.96 & 24.74 & 98.91 & 93.16 &
        86.7 & 80.63 & 68.45 & 60.87 & 99.88 & 81.85 \\
        \end{tabular}
    \caption{Accuracy of evaluating the model without random mask on a dataset
    without (\textsc{ds\oldstylenums{1}}) and with cutoff frequency variations
    (\textsc{ds\oldstylenums{2}}). Varying the cutoff parameter in the codec
    greatly degrades model results. }
    \label{tab:res_exp1b}
\end{table*}

\subsubsection{\textsc{ds\oldstylenums{2}}.} For the second dataset, we use the
same original tracks as were used to create \textsc{ds\oldstylenums{1}}. We also
use the same codec parameters, but vary the cutoff frequency of the codecs,
choosing among 14, 16, 18 and 20 kHz.
\textsc{ds\oldstylenums{1}} and \textsc{ds\oldstylenums{2}}, therefore, differ
only on the lossy versions obtained for each track. We use the same random
70/10/20 split for training/validation/testing for both datasets. All our
experiments are run using \textsc{ds1} for training and validation. Evaluation
is done on \textsc{ds\oldstylenums{1}} (cf.\ Sec.~\ref{sec:experiment1}) or
\textsc{ds\oldstylenums{2}} (Sec.~\ref{sec:experiment2}).

\subsection{Evaluation}
\label{sec:eval}

We evaluate the performance of our lossy/lossless detection model in three ways.
Firstly, we provide quantitative evaluation and report the model accuracy.
Secondly, we inspect saliency maps of the \textsc{cnn} blocks of our model to
gain qualitative insight into what signal properties the model is sensitive to.
Finally, we also inspect the errors of our model in detail to help us assess the
effectiveness our proposed method to make our model more robust, and identify
avenues for future work.

\section{Experiments \& Results}\label{sec:typeset_text}

In this section we first describe our experiments and report our results on a
naively trained lossy/lossless audio detection model (Section
\ref{sec:experiment1}). After an analysis of our results, we report on a more
robust variation of our model in Section \ref{sec:experiment2}, and an analysis
of errors in Section \ref{sec:experiment3}.

\subsection{Experiment 1: Naive Model Training}
\label{sec:experiment1}

In our first experiment, we train our model on \textsc{ds\oldstylenums{1}}. For
each track in our test set, we extract 2-second windows of raw audio with 50\%
overlap. For each window, we perform a forward pass through our trained network,
and collect the output of the classification head. We take the mean of all
windowed local model outputs as the global output per track. 

\subsubsection{Results} In line with previous research (e.g.
\cite{hennequin2017codec, seichter2016aac}), we find near-perfect performance on
lossy/lossless audio detection of audio with default codec settings. The top row
of Table \ref{tab:res_exp1b} shows the results broken down by codec and bit rate
for \textsc{ds\oldstylenums{1}}. We obtain near-perfect results per bit
rate/code combination. On average, we obtain 99.79\% accuracy across all codecs
and lossless files.

However, if we slightly tweak the codec parameters at test time (i.e. we test
our model on \textsc{ds\oldstylenums{2}}) the performance drops significantly.
The bottom row of Table \ref{tab:res_exp1b} shows the results of evaluating the
model on the dataset with cutoff frequency variations. The results show much
poorer results for the lossy tracks across all codec/bit rate combinations.
Specifically, we find a big drop in accuracy of around 70 percentage points for
the \textsc{libfdk\_aac} codec and around 30 percentage points for the
\mpthree{} codec. The \textsc{libvorbis} is less impacted, but is still
significantly impacted by around 10 percentage points.
  
\subsubsection{Analysis}
\label{sec:analysis1}

To get a better sense of what our model has learned, we turn towards a feature
analysis of the \textsc{cnn} part of the network. When inspecting the
spectrogram of a potentially lossy file with the naked eye, one of the most
striking aspects is the nulling of coefficients, resulting in ``holes'' in the
spectrogram. We expected the convolutional part of the network to pick up on
those, and to design features that capture this phenomenon.

\begin{figure}[t]
    \centering
    \includegraphics[width=\columnwidth]{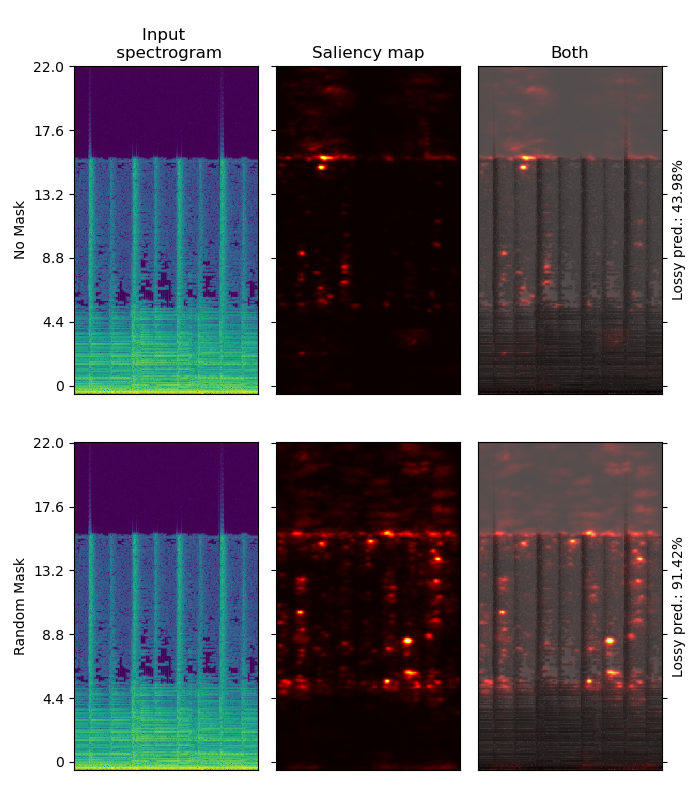}
    \caption{Saliency maps from exposing a model trained without (top) and with
    (bottom) random mask to lossy audio. The model with random mask shows more
    activation in the holes of the spectrogram without losing any of the
    activations at the cutoff frequency. }
    \label{fig:saliencymaps-ds1}
\end{figure}

However, when we visualize saliency maps from our network, we find a different
pattern (see Figure~\ref{fig:saliencymaps-ds1}, top row). 
It seems that the model is more concerned with the cutoff frequency of the lossy 
audio than with the holes in the spectrogram. Although the cutoff frequency is a useful 
feature, by itself it is neither necessary nor sufficient 
to determine whether an audio signal has been encoded with  a lossy codec.

Table \ref{tab:res_exp2} shows the results of the model per cutoff frequency, in
the columns marked with `No'. Here again we see that most cutoff frequency
variations are severely underperforming when compared to the previous test
dataset.

The model performs best at a cutoff frequency of 16 kHz. This can be explained
by the fact that this is the default cutoff frequency of \textsc{libvorbis},
which is therefore not affected by this transformation. In the next section, we
adapt the model to be robust against this cutoff effect.

\begin{table*}[!h]
\centering
\resizebox{\linewidth}{!}{
    \begin{tabular}{l|rrrrrr|rrrrrr|rr}
    
    ~ & 
    \multicolumn{2}{c}{\textsc{libfdk\_aac}} & 
    \multicolumn{2}{c}{\textsc{libvorbis}} & 
    \multicolumn{2}{c}{\textsc{libmp3lame}} &
    \multicolumn{2}{c}{128k} & 
    \multicolumn{2}{c}{256k} & 
    \multicolumn{2}{c}{320k} & 
    \multicolumn{2}{c}{\textsc{mean}} \\
    
    Cutoff & 
    \rotatebox[origin=c]{0}{No} & 
    \rotatebox[origin=c]{0}{Mask} &
    \rotatebox[origin=c]{0}{No} & 
    \rotatebox[origin=c]{0}{Mask} &
    \rotatebox[origin=c]{0}{No} & 
    \rotatebox[origin=c]{0}{Mask} &
    \rotatebox[origin=c]{0}{No} & 
    \rotatebox[origin=c]{0}{Mask} &
    \rotatebox[origin=c]{0}{No} & 
    \rotatebox[origin=c]{0}{Mask} &
    \rotatebox[origin=c]{0}{No} & 
    \rotatebox[origin=c]{0}{Mask} &
    \rotatebox[origin=c]{0}{No} & 
    \rotatebox[origin=c]{0}{Mask} \\

    \hline
    14 kHz & 24.1  & \textbf{81.0} & 100.0 & \textbf{100.0} & 76.9&
    \textbf{100.0} & \textbf{90.4} & 82.6 & 53.7 & \textbf{100.0} & 49.3 &
    \textbf{97.8} & 65.9 & \textbf{93.3} \\
    16 kHz & 83.9  & \textbf{98.4} & 100.0 & \textbf{100.0} & 98.6 &
    \textbf{100.0} & 88.2 & \textbf{100.0} & 97.0 & \textbf{97.7} & 98.6 &
    \textbf{100.0} & 94.6 & \textbf{99.5} \\
    18 kHz & 0.7   & \textbf{86.7} & 66.9 & \textbf{100.0} & 25.3 &
    \textbf{100.0} & 50.0 & \textbf{96.2} & 28.2 & \textbf{94.1} & 12.4 &
    \textbf{94.8} & 29.5 & \textbf{95.6} \\
    20 kHz & 11.1  & \textbf{96.5} & 100.0 & \textbf{100.0} & 82.9 &
    \textbf{100.0} & 46.2 & \textbf{100.0} & 76.1 & \textbf{97.2} & 70.2 &
    \textbf{100.0} & 65.1 & \textbf{98.9} \\
    \hdashline
    \textsc{mean} & 28.3 & \textbf{90.1} & 92.9 & \textbf{100.0} & 70.1 &
    \textbf{100.0} & 70.1 & \textbf{93.6} & 64.1 & \textbf{97.9} & 57.3 &
    \textbf{98.8} & 63.7 & \textbf{96.8} \\
    \end{tabular}
} \caption{Accuracy (in percentage points) of evaluating our models without (No)
and with (Mask) random mask on \textsc{ds\oldstylenums{2}}, per codec and bit
rate, for varying cutoff frequency. Lossless  accuracy is 99.9\% for No and
99.8\% for Mask.}
\label{tab:res_exp2}
\end{table*}


\subsection{Experiment 2: Creating a Robust Model}
\label{sec:experiment2}

In order to increase the model's robustness against the lossy codec's cutoff
frequency, we present a second experiment where we randomly mask the upper end
of the spectrum. The mask, defined in \ref{sec:arch}, is applied to all input
files.

The application of this random mask is intended to force the model not to solely
rely on the codec cutoff frequency to make a prediction, and instead also rely
on other signal degradations included by codecs, such as "holes" in the
spectrogram. A fixed mask at a specific cutoff frequency would have meant
throwing away the information given by the spectral rolloff entirely and this
would have been suboptimal in the opposite direction.


We train this model on \textsc{ds\oldstylenums{1}} and evaluate on
\textsc{ds\oldstylenums{2}}.

\subsubsection{Results} Table \ref{tab:res_exp2} shows the results obtained for
the model trained with the random mask on \textsc{ds\oldstylenums{1}} and
evaluated on \textsc{ds\oldstylenums{2}}. We observe good classification results
on average, 96.8\% on lossy files and 99.8\% on lossless files. Overall, we
obtain 98.4\% lossy/lossless classification accuracy across the entire test
dataset. Comparing with the naive model, the accuracy on
\textsc{ds\oldstylenums{2}} improves significantly across the board. 

With the mean classification accuracy at 90\% or above in all conditions (last
column of the table), this model is broadly robust against cutoff frequency
variations. It is interesting to note that performance on the AAC codec is
comparatively lower than on other codecs. This result suggests that the AAC
codec is more challenging to detect, and warrants further investigation, which
we leave for future work. We hypothesise it may be due to the AAC codec
producing less artefacts in the magnitude spectrogram. 

\begin{figure}[t]
\minipage{0.49\columnwidth}
  \includegraphics[width=\linewidth]{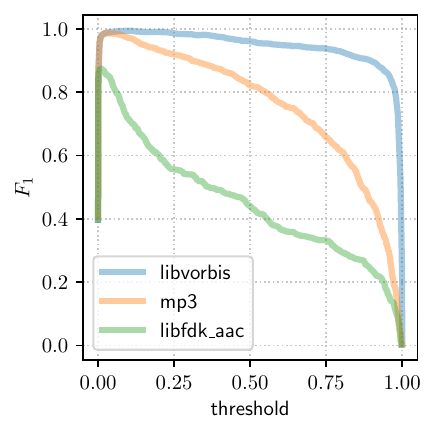}
\endminipage\hfill
\minipage{0.49\columnwidth}
  \includegraphics[width=\linewidth]{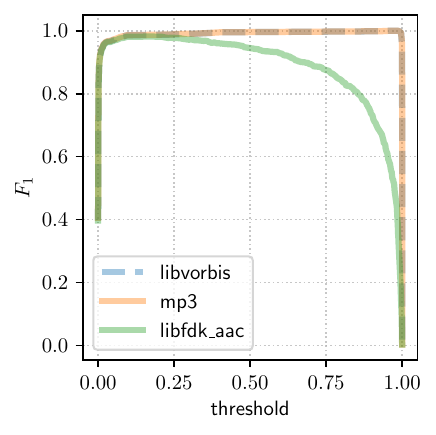}
\endminipage\hfill \caption{F1-score for varying thresholds, 
evaluated on \textsc{ds\oldstylenums{2}}. Each line analyses
the subset made of lossless files as negatives and the specified codec as
positives; files encoded with different codecs are discarded. Left: model
without random mask; Right: model with random mask.}
\label{fig:F1-score}
\end{figure}

In Fig. \ref{fig:F1-score}, for both the model without random mask (cf.
Section~\ref{sec:experiment1}) and the model with (cf.
Section~\ref{sec:experiment2}) we plot the $F_1$ score (i.e., the harmonic mean
between precision and recall) as a function of the threshold of the binary
classification prediction. The $F_1$-score for the model without the random mask
peaks at very low values of the threshold and then decays for increasing
threshold at a rate that highly depends on the codec analysed.

This suggests three conclusions: (1) There are a number of test set files that
yield a prediction $p(x)$ in the central region $0.1 < p(x) < 0.9$, which shows
a high degree of uncertainty for the model; (2) since the F1-score is
monotonously decreasing, the model tends to output false negatives rather than
false positives; (3) different codecs are identified with different level of
proficiency.

Compare this with the output for the model with the random mask: in the case of
\textsc{libmp3lame} and \textsc{libvorbis} codecs, the $F_1$-score is almost
flat and close to 1 for the entire range of thresholds. The \textsc{libfdk\_aac}
codec still shows some decrease in performance for increasing thresholds, but
the peak value increased from 0.875 to 0.982 and the area under the $F_1$ curve
jumped from 0.450 to 0.891. From the results above we can conclude that the
introduction of the random mask brings higher peak performance and also reduces
the impact of the choice of the threshold.

\subsubsection{Analysis}
Similarly to the analysis presented in Section~\ref{sec:analysis1}, we visualise
saliency maps of the model trained with the random mask in the bottom row of
Figure~\ref{fig:saliencymaps-ds1}. Compared to the saliency of the model with no
mask (top row), we see a much brighter activation in the holes of the
spectrogram without losing any of the activations at the cutoff frequency. The
model has learned to rely on more markers to make its choice.

\subsection{Qualitative Analysis of Errors}
\label{sec:experiment3}

In this section, we present a qualitative analysis of the erroneous predictions
produced by our model trained with the random mask.

\subsubsection{Lossless Errors.}
\label{sec:errors}
\begin{figure}[t]
    \centering
    \includegraphics[width=\columnwidth]{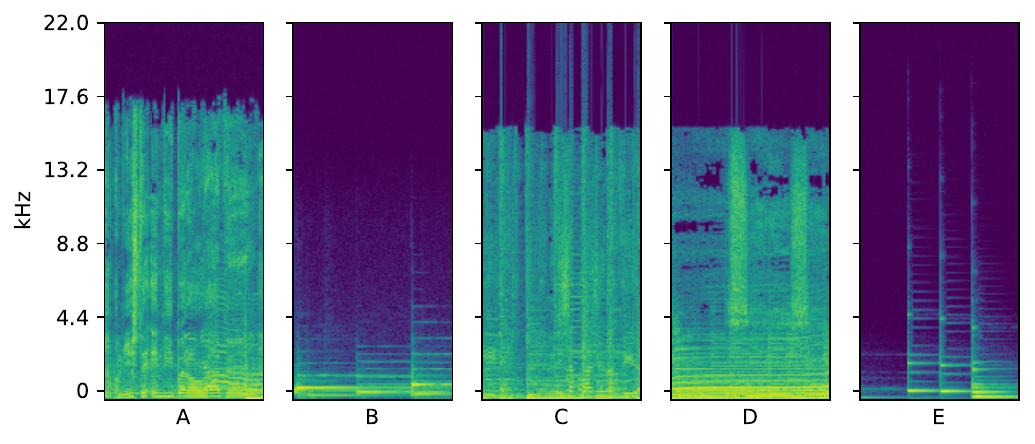}
    \caption{The five assumed lossless tracks misidentified
    as lossy. However, A, C, and D are in fact lossy. 
    B and E are quiet tracks with a single instrument.}
    \label{fig:lossless errors}
\end{figure}

From our entire test subset of \textsc{ds}\oldstylenums{2}, we observe only 5
cases (0.2\%) where the model made a "lossy" prediction while the recording is
in the lossless part of the dataset.  
The spectrogram of three out of the five tracks (A, C and D in 
Figure \ref{fig:lossless errors}) show the hallmarks
of lossy compression. It appears that our model was indeed correct in predicting
a lossy encoding, and therefore revealed "in-the-wild" cases of accidental lossy
compression that were present in our dataset. 

The other two tracks (B and E) are quiet and sparsely orchestrated tracks. It is
notable that the spectrogram also appears sparse, with very little energy in the
upper frequency range. Given that lossy codecs often feature energy depletion in
the top part of the frequency range, we hypothesise that the misclassification
may be due to the model relying on the absence of energy in the upper register
in this case. 

\subsubsection{Lossy Errors.}
\begin{figure}[t]
    \centering
    \includegraphics[width=\columnwidth]{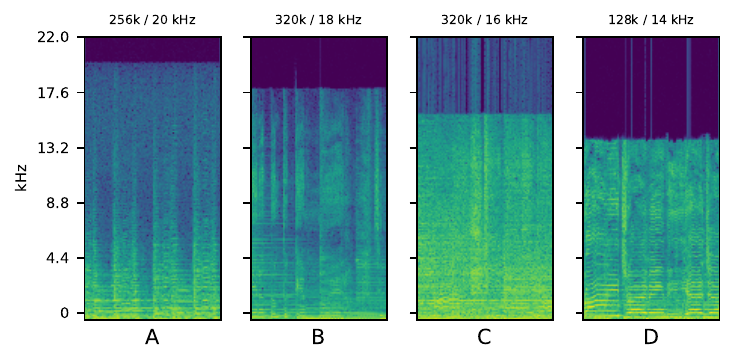}
    \caption{A random selection of lossy tracks misidentified as lossless. All
    tracks are encodec with \textsc{libfdk\_aac}. The spectrograms show less
    holes and band rupture compared to other codecs, especially under 14 kHz.}
    \label{fig:lossy errors}
\end{figure}

Table \ref{tab:res_exp2} shows that the entirety of cases where the model
erroneously classified recordings as lossless when it should be lossy comes from
tracks encoded with the \textsc{libfdk\_aac} codec. 
In Figure \ref{fig:lossy errors}, spectrograms of 2 second excerpts from a
random selection of error tracks are visualized. 

From inspecting the spectrograms of \textsc{libfdk\_aac} encoded tracks, we find
that common characteristics are (1) the spectral roll-off is relatively stable
over time, (2) the preservation of transients above the cutoff frequency, which
can often span upwards to the Nyquist frequency, and (3) less nulling of
spectral coefficients, resulting in fewer holes in the spectrogram. The
\textsc{libfdk\_aac} codec is a superior codec in terms of compression
efficiency, meaning it can provide better audio quality at lower bitrates than
other codecs \cite{brandenburg1999mp3}. 

Table 2 shows that AAC with cutoff 14kHz is only 81\% accuracy. We hypothesize
that the \textsc{libfdk\_aac} codec does not produce as much ``holes'' in the
spectrogram below this threshold. Our model applies the random mask to every
example in our training dataset, which can be confusing on \textsc{libfdk\_aac}
samples. That is, as the random mask is applied at a relatively low cutoff
frequency, the resulting spectrogram is almost identical to a lossless example.
One avenue for future work could be to apply the random mask with a lower
probability, to allow the model to also learn other spectral characteristics of
\textsc{libfdk\_aac} samples.

\section{Conclusion}

In this paper, we presented a lossy audio compression detection method that can
robustly estimate whether a given audio file has been lossy encoded before. We
show that naively training a model  results in near-perfect lossy audio
compression detection on the held-out test set generated using the same encoding
parameters. 

However, we find that, for several widely used lossy codecs, the performance of
this model catastrophically degrades when exposed to variations of the cutoff
frequency parameter that were not seen during training. This result suggests
that a naively trained model is overly reliant on the cutoff value. In response
to this shortcoming, we propose to amend the training strategy by applying a
random mask to the upper range of the spectrogram, in order to reduce the
model's reliance on the codec cutoff frequency value. 

We show that this method results in a model that is significantly more robust
against frequency cutoff variations. Our experiments reveal compelling
performance on all codec and bit rate combinations we considered, but reveal
that there remains room for improvement on the detection of the
\textsc{libfdk\_aac} codec. We hypothesise that the \textsc{aac} codec is
comparatively more difficult to detect than \mpthree{} and Ogg Vorbis because it
generates less artefacts in the magnitude spectrogram.  An avenue for future
work may consist in exploring further development of the training strategy in
order to improve performance on the \textsc{aac} codec. 

\bibliography{ISMIRtemplate}

\end{document}